\documentclass[12pt]{iopart}
\usepackage{epsfig}
\usepackage{graphicx}

\begin{document} 
\title[Crackling noise during failure of alumina]{Crackling noise during failure of alumina under compression: effect of porosity} 

\author{Pedro O Castillo-Villa$^1$, Jordi Bar\'o$^1$, Antoni Planes$^1$, Ekhard K H Salje$^2$, Pathikumar Sellappan$^3$, Waltraud M Kriven$^3$ and Eduard Vives$^1$} 
\address{$^1$ Departament d'Estructura i Constituents de la Mat\`eria,
Facultat de F\'{\i}sica, Universitat de Barcelona, Mart\'{\i} i
Franqu\`{e}s 1, E-08028 Barcelona, Catalonia, Spain.}
\address{$^2$ Department of Earth Sciences, University of Cambridge,
Downing Street, Cambridge CB2 3EQ, UK.}
\address{$^3$ Department of Materials Science and Engineering, University of Illinois at Urbana-Champaign, USA.}
%
%
%
%
%
%
%
%
%
%
%
%
%
%
%
 \begin{abstract}
We study the acoustic emission avalanches during the failure process of porous alumina samples (Al$_2$O$_3$) under compression. Specimens with different porosities ranging from 30\% to 59\% have been synthetized from a mixture of fine-grained alumina and graphite. The compressive strength as well as the characteristics of the acoustic activity have been determined. The statistical analysis of the recorded acoustic emission pulses reveals, for all porosities, a broad distribution of energies with a fat tail, compatible with the existence of an underlying critical point. In the region of 35\%-55\% porosity, the energy distributions of the acoustic emission signals are compatible with a power law behaviour over two decades in energy with an exponent $\epsilon = 1.8 \pm0.1$.
\end{abstract}


\submitto{Journal of Physics: Condensed Matter}

\maketitle

\section{Introduction}
Understanding the failure of materials under compression has important implications for the prediction of collapses in many natural and man-made structures: mines, buildings, bones, etc.  At very large scales, even Earthquakes can be regarded as produced by failures of the Earth crust under compressive stresses. In these examples, the compressed materials exhibit a high degree of porosity which, in many cases, consists in voids with a complex geometry with sizes that range from mm to nm \cite{Dunlop2013}. The problem has, therefore, a multiscale nature which makes it very difficult to formulate quantitative models \cite{Girard2012}.

Macroscopic observations show that the compression of porous materials leads to sudden partial collapses rather than a smooth elastic or plastic deformation. The partial collapses are named ‘jerks’ and, within the context of out-of-equilibrium dynamics have been classified as avalanche phenomena. Their  statistical features are similar to crackling noise as reviewed by Sethna et al \cite{Sethna2001}. Jerks are also seen, not only under compression, but also under shear deformations where the microstructures of the sample change in sudden movements rather than continuously \cite{Romero2011,Harrison2010,Salje2009}. 

One of the experimental techniques that has been succesful in understanding  the statistical properties of jerks is the detection of the Acoustic Emission (AE) associated with the microcracks ocurring in the samples. This experimental technique already revealed strong similarities between the compression of natural rocks with low porosities and earthquakes, more than 5 years ago \cite{Davidsen2007}. Very recent measurements\cite{Salje2011a,Baro2013} on the compression of synthetic porous SiO$_2$ (Vycor) which has a 40\% porosity  have shown that the event energies and the times of occurrence show strong similarities with earthquake statistics, both concerning temporal correlations (aftershocks) and energy distributions. 
In particular, the distribution of event energies shows a Gutenberg-Richter behaviour. i.e. a lack of charactersitic scales: the probability of a jerk with energy $E$ follows $P(E) \sim  E^{-\epsilon}$  with $\epsilon=1.40\pm0.05$.  The energy interval for the power-law behaviour in this experiment \cite{Salje2011a,Baro2012} spans over more than 4 decades.

Similar measurements in not so well defined materials like the mineral goethite have also been performed\cite{Salje2013}. They lead to slightly higher exponents and a more limited energy interval in which the power-law decay can be measured. Goethite data shows an increase of the energy exponent with increasing porosity between $\epsilon=1.60\pm0.05$ and $\epsilon=2.0\pm0.1$ in the porosity range between 55\% and 89\% \cite{Salje2013}. 

Therefore, at present there is no clear understanding about what factors may alter the energy distribution of AE events. Interesting questions are: Is the degree of porosity a relevant parameter determining the critical properties? Is there a unique porosity value for which the distribution of energies is really critical? What is the influcence of the material properties?
 
The two materials studied so far (Vycor and natural goethite samples) are relatively soft. This raises the question whether much harder materials follow the same trend. This question is important also for the potential application of the results for  the analysis of earthquakes. Faulting during earthquakes evolves from the sliding of minerals against each other.  These movements involve the breaking of chemical bonds. These bonds are, in the earth's crust, in their majority related to Si-O,  Al-O  and Fe-O, so that the sliding mechanism becomes  akin to the collapse of porous minerals where similar phenomena determine locally the collapse mechanism.   We have investigated the Si-O bond breaking in the synthetic SiO$_2$ material, "Vycor" and the Fe-O bond-breaking in the natural mineral goethite.  In this paper we will show that the Al-O bond breaking in synthetic alumina ceramics leads to very similar jerk distributions as in the other samples. We focus in this study on alumina, Al$_2$O$_3$. One of its crystalline polymorphic phases, $\alpha$-Al$_2$O$_3$ is corundum and the only thermally stable, equilibrium phase. Diamond does not form significant porosity so that corundum becomes a materials of choice for this study. 

Comparing the three materials with vastly different bond strengths we will show that there is no systematic dependence on the bond strength. There is some indication that the degree of porosity does play a minor role in the effective energy exponent, however. We will argue that this dependence relates to the criticality of the collapse process and the effect of the cut-off of a limited range of power law dependences rather than as an intrinsic feature of the topology of the pore distribution of the porous materials.

\section{Experimental}

\subsection{Synthesis and characterization of the samples}

Porous and phase-pure, alumina samples were produced using commercially available alumina (Almatis, Leetsdale, PA, USA) and graphite powders (Aldrich Chemical Company, St. Louis, MO, USA) as precursor materials. The fine alumina powders were better than 99.8\% chemically pure with an average particle size of (D$_{50}$) 0.45 $\mu$m, surface area of 8.5 m$^2$/g and had a density of 3.98$\pm$0.01 g/cm$^3$. Graphite particles were used as pore formers and had 1.9 g/cm$^3$ density and 1-2 $\mu$m particle size. Appropriate amounts of graphite particles, namely 20, 30, 40, 50, 60, and 70 vol\% were introduced into the alumina powder matrix to prepare alumina-graphite mixtures by milling in a ball mill (at 100 rpm for 24 h). 1 wt\% of polyethylene glycol (PEG, Mn = 200, Sigma Aldrich, St. Louis, MO, USA) also added to the mixture as a binder and ethanol (99.5 \%, Acros Organics, Geel, Belgium) was used as a solvent with yttria stabilized zirconia cylinders as milling media. The ball milled slurry was then dried using a hot plate, while continuously stirring to remove the ethanol and then dried at ~ 100 $^\circ$C for 24 h and stored.
The dried and crushed powders were initially compacted using a 19 mm cylindrical hardened steel die, uniaxially pressed under  less than 5 MPa and cold isostatically pressed (CIP) under $\sim$ 344 MPa  for 10 minutes. A cold isostatic press (CIP, Model CP 360, American Isostatic press, Columbus, OH, USA) was employed to consolidate the powder particles homogeneously and the high pressure involved also allowed removal of intergranular pores and cracks, which would result in inhomogenity in the bulk samples during and after sintering. CIPed samples were then heated slowly to 900 $^\circ$C at a heating rate of 1 $^\circ$C/min to remove the pore formers (graphite particles) and binders and then heated to 1450 $^\circ$C at a heating rate of 5 $^\circ$C/min and held for 3 h. Powder X-ray diffractometry (XRD) with Cu K$\alpha$ radiation (Siemens-Bruker D5000, Germany) was used to analyze the phases present in the materials.

The average bulk densities and apparent porosity values were measured by the Archimedes method using boiling water (ASTM C373).  Table \ref{TAB1} lists the density and apparent porosity of the studied samples.

\begin{table}[htb]
\begin{center}
\begin{tabular}{|c|c|c|c|}
\hline
Sample & Graphite & Bulk density & Apparent porosity\\
 & \% vol & g/cm$^3$ & \% \\
\hline
Alumina20 & 20 & 3.15 $\pm$ 0.01	& 21.2 $\pm$ 0.06  \\
Alumina30 & 30 & 2.78 $\pm$ 0.01  &	30.3 $\pm$ 0.04  \\
Alumina40 & 40 & 2.52 $\pm$ 0.01 & 	36.8 $\pm$ 0.06 \\
Alumina50 & 50 & 2.16 $\pm$ 0.01 & 45.8 $\pm$ 0.04 \\
Alumina60 & 60 & 1.91 $\pm$ 0.01	 & 52.2 $\pm$ 0.05 \\
Alumina70 & 70 & 1.66 $\pm$ 0.01	& 58.5 $\pm$ 0.01\\
\hline
\end{tabular}
\end{center}
\caption{\label{TAB1} Density and apparent porosity for porous alumina samples prepared using A16SG alumina powders and graphite particles. Apparent porosity values were calculated using bulk density values and a theoretical density value of alumina (3.98 $\pm$ 0.01 g/cm$^3$).}
\end{table}

\subsection{Phase and microstructural analysis.}

\begin{figure}
\begin{center}
\epsfig{file=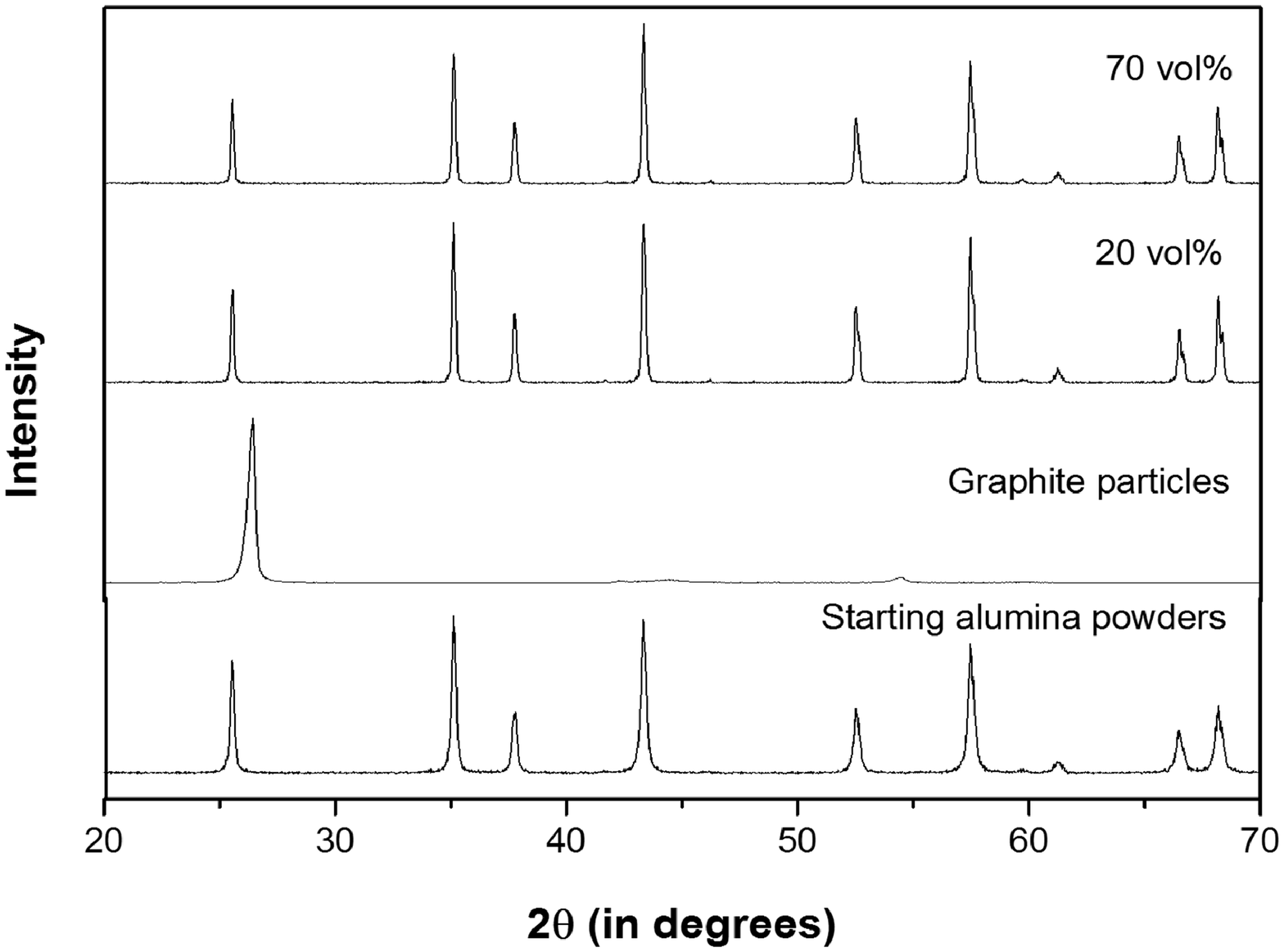,width=6.5cm,angle=270,clip=}
 \end{center} 
\caption{\label{FIG1} XRD diffractometry patterns of porous alumina samples prepared using 20 and 70 vol\% graphite particles dispersed in alumina particles, compared with the starting alumina and graphite particles.}
\end{figure}

In order to analyze the phases present in the processed porous alumina pellets, XRD patterns of the 20 vol\% (lowest pore former values) and 70 vol\% (highest pore former values) were compared with the XRD patterns of the starting alumina and graphite particles. Results are shown in Fig.~\ref{FIG1}. Adding the graphite particles to alumina and the processing conditions did not result in any intermediate phase formation, even though the graphite particle content increased from 20 to 70 vol\%.  XRD analysis confirmed that the pore former phase (graphite particles) was completely removed during the heat treatment.

\begin{figure*}
\begin{center}
\epsfig{file=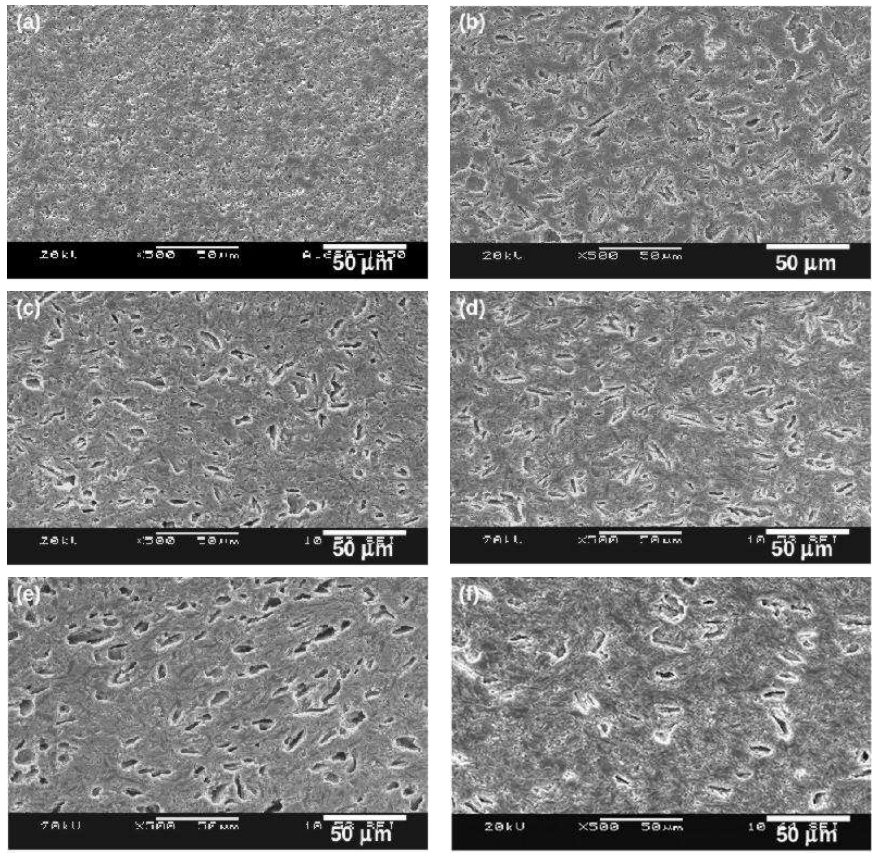,width=12cm,clip=}
 \end{center} 
\caption{\label{FIG2} SEM micrographs of polished alumina and porous alumina samples sintered at 1450 $^\circ$C for 3 h. (a) Alumina without any pore formers, (b) 20 vol\% graphite particle content, (c) 30 vol\% graphite particle content,(d) 40 vol\% graphite particle content,(e) 50 vol\% graphite particle content, and (f) 60 vol\% graphite particle content. All images were observed in the normal SEM image mode under the same conditions at 500$\times$ magnification.}
\end{figure*}

\begin{figure}
\begin{center}
\epsfig{file=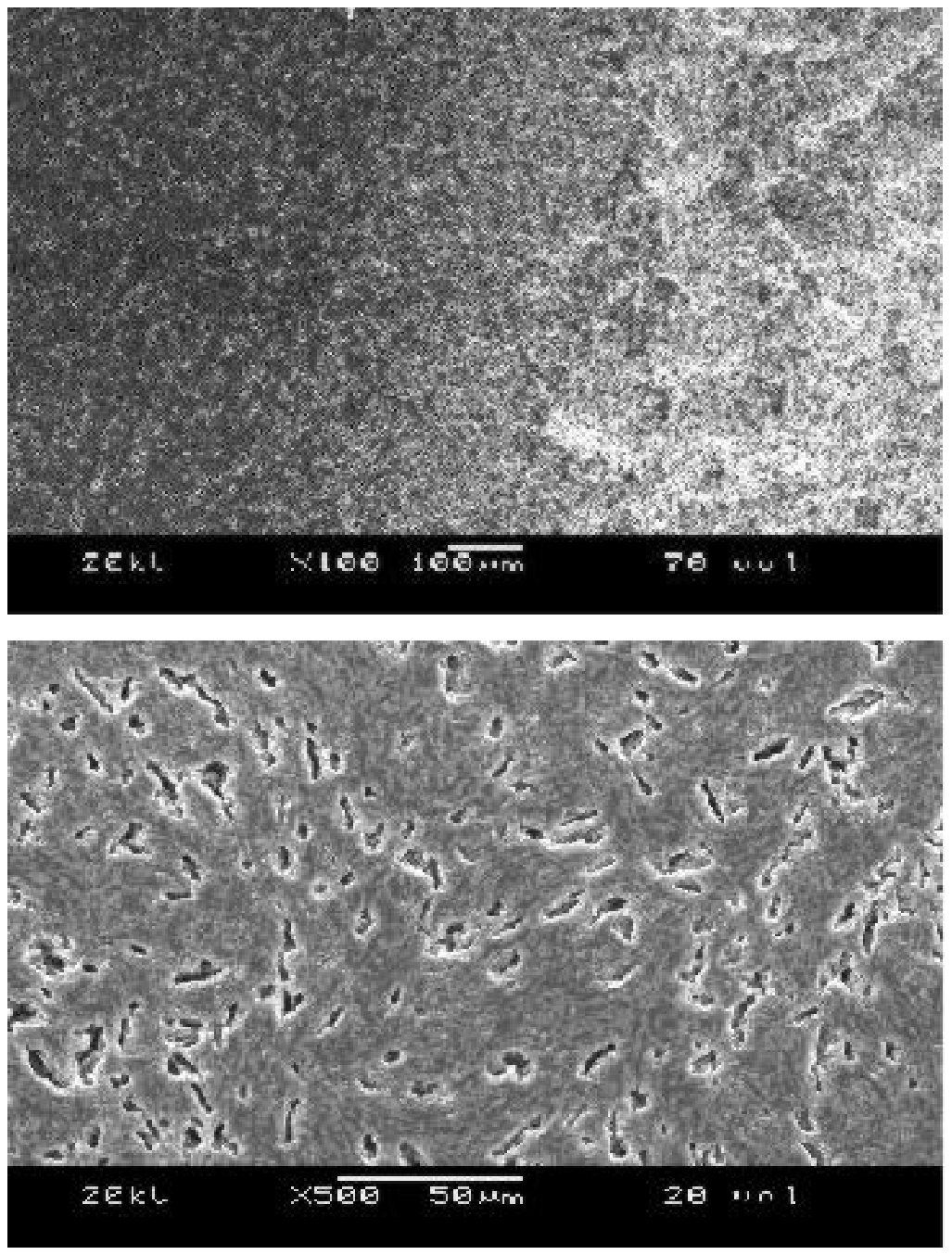,width=7.5cm,clip=}
 \end{center} 
\caption{\label{FIG3} SEM micrographs of polished alumina-70 vol\% graphite particle content sample sintered at 1450 $^\circ$C for 3 h. }
\end{figure}

Samples were sectioned using a low speed diamond tipped saw and cross-sectioned regions were polished down to 0.25 $\mu$m to reveal the microstructure. Fig.~\ref{FIG2}(a) shows the microstructure of sintered alumina samples with uniformly distributed fine porosity (pore sizes were $<$ 5 $\mu$m). Very fine pores are a common feature in alumina microstructures when sintering takes place without high pressure or sintering aids \cite{Rahaman2003}. In the present case, the fine particle size of the starting material, high surface area and high cold isostatic compaction resulted in $\sim $ 97\% dense material compared to the theoretical density of pure $\alpha$-alumina. The microstructures from Figs.~\ref{FIG2}(b) to Fig.~\ref{FIG2}(e) show that the porosity level increased linearly as the graphite particles content (pore formers) increased from 0 to 50 vol\% as compared to Fig.~\ref{FIG2}(a).
The pores were very fine (less than or equal to 10 $\mu$m) and uniformly distributed on the alumina matrix. Fig.~\ref{FIG2}(f) shows that for 60 vol\% graphite content, the pore distribution is less uniform compared with less porous samples. This trend was more severe when the graphite particle content increased to nominally 70 vol\% as shown in Figs.~\ref{FIG3}(a) and \ref{FIG3}(b).

\subsection{Compression and Acoustic Emission Setups}

The experimental arrangement for the uniaxial compressional test has been described elsewhere\cite{Salje2011a}. It consists of two parallel circular aluminium plates perpendicular to the vertical direction. The bottom plate, hanging from the load cell at the top of the arrangement, is static. The upper plate is pulled downwards by means of three guides sliding through convenient holes drilled in the bottom plate. The pulling device consists of a water container acting as a dead load. Small  pump rates for the inflowing water allowed to impose a slowly increasing load. Acoustic emission sensors are embedded into the compression plates, centered at a distance of 4 mm from the sample surface. The sensors are acoustically coupled to the aluminium plates by vaseline. A small amount of vaseline is also placed between the samples and the aluminum plates. The signal from the sensors is preamplified to 60dB and input in a PCI-2 system (Europhysical Acoustics, Mistras group, France) operating at  10 MHz. 

A laser extensometer (Fiedler Optoelektronik, Germany) measures the vertical separation between the plates with a resolution of 100 nm. The load cell (1 kN range) signal is read with a lock-in amplifier and has been calibrated with standard weights.

Special care has been taken in order to avoid noise from friction of the guides. By performing blank measurements, we have identified the properties of the noise and have designed software filters in order to supress such signals from the statistical analysis that will be presented.

In order to perform the compression experiments, the original cylindrical samples have been cut using a stainless-steel blade into suitable parallelepipedic specimens. For high porosities, the specimens tend to be more irregular and acceptable sample shapes are only obtained after some trial and error. Details of the specimens studied are presented in Table \ref{TAB2}. Height of the specimens ($z$) has been choosen in the range 3-6 mm, which is suitable for the laser extensometer range. Transverse sections ($A$) have been choosen in order to have failure strengths below the maximum load  that can be applied in our setup (900 N). In order to do systematic comparisons, we have chosen the transverse sections of a series of specimens (2A30, 2A40,.. 2A70) with different porosities to have similar nominal compression rates in the range 22-47 kPa/s. Moreover, for the samples with porosities  36.8\% and 52.2\%, specimens with  different transversal sections (F401-F407, F601-F606) have been choosen in order to perform a study as a function of the compression rate in the range 5kPa/s - 80 kPa/s.

\begin{table*}[htb]
\begin{center}
\begin{tabular}{|l|c|c|c|c|c|c|}
\hline
Specimen & App. Porosity & Height & Section & Mass & Stress rate & Comp. Strength \\
& (\%) & (mm) & (mm$^2$) & (mg) & (kPa/s) & (MPa) \\
\hline
A30 & 30.3 $\pm$ 0.04 & 3.75 & 5.04 & 48 & 27.40 & 188.07 \\
2A30 & 30.3 $\pm$ 0.04 & 4.05 & 3.90 & 41 & 34.80 & 221.39 \\
A40 & 36.8 $\pm$ 0.06 & 4.35 & 7.45 & 72 & 18.80 & 73.84 \\
2A40 & 36.8 $\pm$ 0.06 & 4.50 & 6.00 & 60 & 22.00 & 92.70 \\
F401 & 36.8 $\pm$ 0.06 & 4.35 & 3.96 & 35 & 11.90 & 91.20 \\
F402 & 36.8 $\pm$ 0.06 & 4.10 & 4.00 & 38 & 25.72 & 93.60 \\
F403 & 36.8 $\pm$ 0.06 & 4.20 & 4.09 & 35 & 33.75 & 87.20 \\
F404 & 36.8 $\pm$ 0.06 & 4.05 & 4.00 & 35 & 45.72 & 80.50 \\
F405 & 36.8 $\pm$ 0.06 & 3.90 & 3.71 & 29 & 63.93 & 94.89 \\
F407 & 36.8 $\pm$ 0.06 & 3.75 & 4.20 & 31 & 80.43 & 86.70 \\
A50 & 45.8 $\pm$ 0.04 & 4.65 & 15.44 & 142 & 28.60 & 51.55 \\
A50DF & 45.8 $\pm$ 0.04 & 4.75 & 15.41 & 143 & 8.00 & 51.11 \\
2A50 & 45.8 $\pm$ 0.04 & 5.25 & 5.75 & 62 & 25.10 & 75.41 \\
2A50R & 45.8 $\pm$ 0.04 & 4.80 & 5.40 & 51 & 41.10 & 97.98 \\
A60 & 52.2 $\pm$ 0.05 & 5.45 & 12.48 & 119 & 12.20 & 54.74 \\
2A60 & 52.2 $\pm$ 0.05 & 4.85 & 5.06 & 41 & 29.20 & 49.29 \\
F601 & 52.2 $\pm$ 0.05 & 4.00 & 5.06 & 35 & 5.63 & 63.96 \\
F602 & 52.2 $\pm$ 0.05 & 4.40 & 5.18 & 42 & 17.05 & 56.85 \\
F603 & 52.2 $\pm$ 0.05 & 4.50 & 5.18 & 42 & 24.57 & 64.02 \\
F604 & 52.2 $\pm$ 0.05 & 3.90 & 5.18 & 37 & 34.44 & 67.48 \\
F604R & 52.2 $\pm$ 0.05 & 3.20 & 5.00 & 25 & 37.03 & 83.14 \\
F605 & 52.2 $\pm$ 0.05 & 4.65 & 4.73 & 36 & 47.96 & 70.58 \\
F606 & 52.2 $\pm$ 0.05 & 4.90 & 4.73 & 40 & 67.10 & 70.05 \\
A70 & 58.8 $\pm$ 0.01 & 5.45 & 13.32 & 97 & 12.20 & 20.49 \\
2A70 & 58.8 $\pm$ 0.01 & 4.95 & 6.37 & 47 & 22.20 & 29.92 \\
\hline
\end{tabular}
\end{center}
\caption{\label{TAB2} Summary of the dimensions, compression conditions and neasured compression strength for each studied specimen.}
\end{table*}

\section{Analysis of acoustic activity}
The analysis of acoustic emission  reveals that the failure process under compression  is rather complex, exhibiting different regimes that can be distinguished by  the qualitative 
measurement of the activity rate (number of signals detected per unit time), the behaviour of the sample height as a function to time $z(t)$ and the square ot the velocity $(dz/dt)^2$. 
Some illustrative examples are shown in Fig. \ref{FIG4}, corresponding to different porosities and similar nominal compressions in the range 22-47 kPa/s. Note the logarithmic scale for the AE activity and the square velocity. One can identify the following regimes:

\begin{figure*}[htb] 
\begin{center}
\epsfig{file=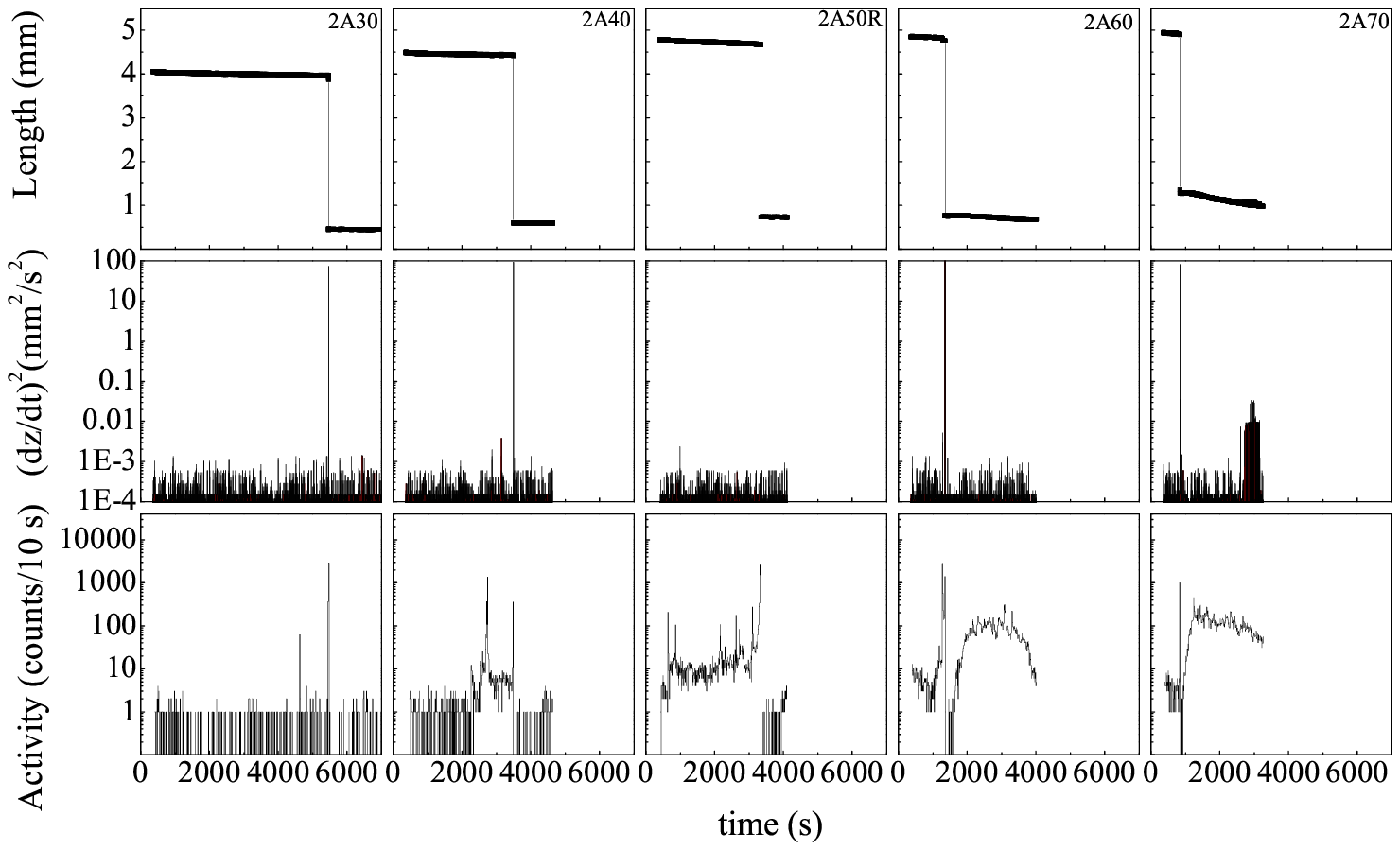,width=14cm,clip=}
 \end{center} 
\caption{\label{FIG4} Examples of compression experiments at $32 \pm 10$ kPa/s. Different columns correspond to different specimens. The first row shows the behavior of the sample height, the second row the behaviour of the square velocity and the third row the AE activity.}
\end{figure*}

\begin{enumerate}

\item {\bf Adaptation}: After the sample is placed between the compression plates, we have waited $\sim$200 s without increasing the load in order to stabilize oscillations of the compressing setup. This first regime shows already some acoustic activity since the compression plates (and the empty container) already represent a certain load on the sample.  It was not further considered for the analysis because it was too difficult to separate the instrumental noise from the signals related to the collapse of the sample.

\item  {\bf Loading}: The second regime starts with the onset of  a constant compression rate and ends after a big failure event or crash. These big failure events reduce the sample height by an amount $\Delta z$,  representing more than a  50\% relative variation ($> 3$ mm). The loading regime is not fully stationary: there are sudden steps in $z$ and/or short transient periods with peaks of acoustic activity (see Fig.~\ref{FIG5}). The statistical variability of the AE signal is very large in this regime. In some cases we observe a first run with stationary low activity while the repetition of the experiment with a different specimen differed greatly. The time evolution of the AE signals were poorly reproducible which resembles observations of seismicity during earth quakes. The value of the applied force at the start of the major failure event ($F_c$) defines the compressive strength $P_c = F_c/A$ where $A$ is the sample transverse section measured at the beginning of the experiment. The excellent time resolution of our experiment allowed us to measure the duration $\Delta t$ of the big failure. During this crash we find a multitude of individual AE events so that the major event is not a ‘snapping’ of the sample, but rather is a jerky avalanche event in its own right. The statistical properties of the signals during the crash are not very different from those found in the whole loading regime. In particular, the signals during the crash are not necessarily more energetic than those previous to the crash. An illustrative example is shown in Fig.~\ref{FIG5} that corresponds to an enlarged section of the experiment 2A60 shown in Fig.~\ref{FIG4}. The dots indicate the detection of AE events and the color the energy of each event. These figures show that large precursor events occur well before the major failure event. After the crash, the sample height reaches again a very stable value that can be identified with a sharp change of the slope in the $z(t)$ plot  (see Fig.\ref{FIG4}). We can define an average failure speed $v = \Delta z/ \Delta t$. The duration $\Delta t$ of the big failure  correlates with porosity: highly porous samples show a shorter crash duration.

\begin{figure}[htb] 
\begin{center}
\epsfig{file=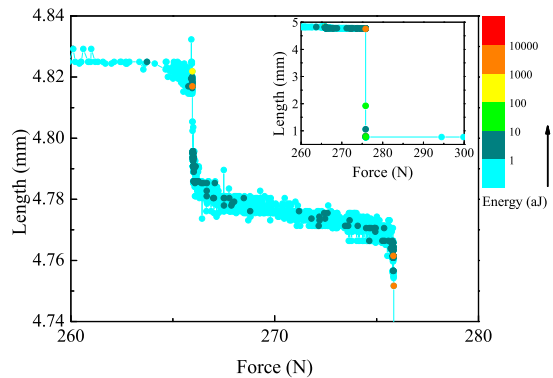,width=7.5cm,clip=}
 \end{center} 
\caption{\label{FIG5} Detailed failure of a sample with 52.2\% porosity. Each data point corresponds to an individual AE signal and the color indicates its energy. The insets show the full scale plot.}
\end{figure}

\item {\bf Pause}: The major failure event is followed by a long period with no AE signals (activity below 1 count/s in our experiment). This regime lasts longer for higher porosities. We will argue that the duration of the pause $\Delta t_{pause}$ relates to the momentum absorbed by the sample at the end of the big failure event.

\item {\bf Fragmentation}:  When  the compression load increases further, after the pause, we observe a fourth regime with rather stationary activity. We associate this regime with the consecutive fragmentation of the pieces of the material that are left after the major failure. Within the durations of the experiments presented here (several hours) and at the studied compression rates, we have not found evidences of how this regime ends.

\end{enumerate}

\begin{figure}[htb]
\begin{center}
\epsfig{file=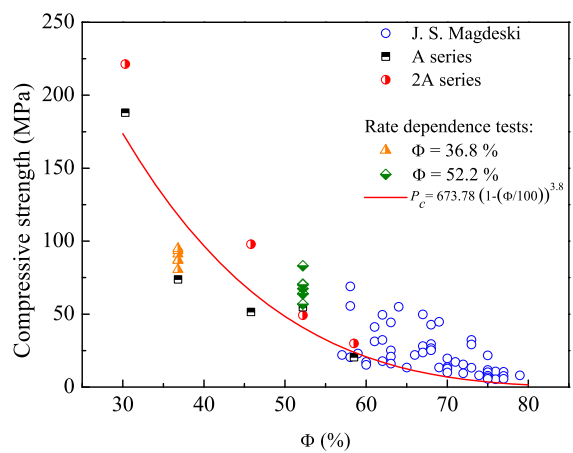,width=7.5cm,clip=}
\end{center} 
\caption{\label{FIG6} Compressive strength as a function of porosity. Data from this work is compared with data in the bibliography. 
The line shows a fit of the behaviour $(1-\Phi/100)^m$, suggested in Ref.~\cite{Salje2010}.}
\end{figure}

Fig.~\ref{FIG6} shows the behavior of the compressive strength ($P_c$) as a function of the porosity ($\Phi$). The observed behavior is fully compatible with the measurements of Magdeski\cite{Magdeski2010}. The line shows the fit of the behaviour $P_c \propto [1 - \Phi/100 ]^m$ with $m=3.8$. This value is in agreement with experimental findings  \cite{Salje2010} but much larger than the predictions based on isotropic reticular models \cite{Liu2010}.

\begin{figure}[htb] 
\begin{center}
\epsfig{file=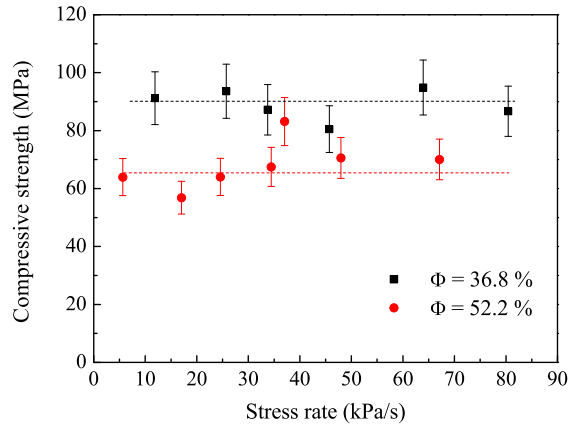,width=7.5cm,clip=}
\end{center} 
\caption{\label{FIG7} Compressive strength as a function of the compression rate for samples with 52.2\%  and 36.8 \% porosity.  Error bars include the differences found between repetitions of the experiment.
} 
 \end{figure}

We measured, in more detail, the collapse of samples with 52.2\% and 36.8\% porosity as function of the compression rate in the range from 5 kPa/s to 70 kPa/s. As illustrated in Fig.~\ref{FIG7}, no significant dependence of the compressive strength $P_c$ on the rate was found.

\begin{figure}[htb]
\begin{center}
\epsfig{file=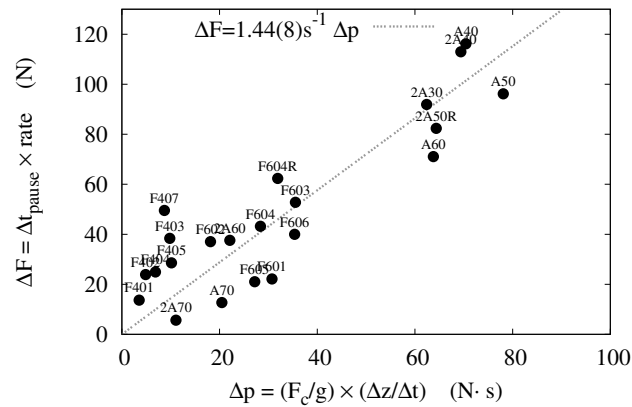,width=6cm,angle=270,clip=}
\end{center} 
\caption{\label{FIG8} Duration of the pause regime (expressed as a force increment) as a function of the impulse delivered to the sample by the fallling weight during the big failure event. Labels indicate the specimen from Table II.}
 \end{figure}

The study of the failure details allows us to extract some conclusions about the origin of the pause regime. Fig.~\ref{FIG8} shows the duration of the pause, expressed as a force interval ($\Delta F$) as a function of the impulse ($\Delta p$) transmitted to the sample when the big crash is arrested. This impulse can be estimated by multiplying the falling mass at the failure point 
($F_c/g$) by the average speed ($\Delta z/ \Delta t$) during the crash. The figure reveals a clear correlation between the two quantities. We argue that the pause in the AE activity is, therefore, due to an effective overshoot associated with this momentun tranfer ($\Delta p$) that deactivates all the avalanches that would have ocurred within $F_c$ and $F_c+\Delta F$. 

\section{Energy distributions}

\begin{figure}[htb] 
\begin{center}
\epsfig{file=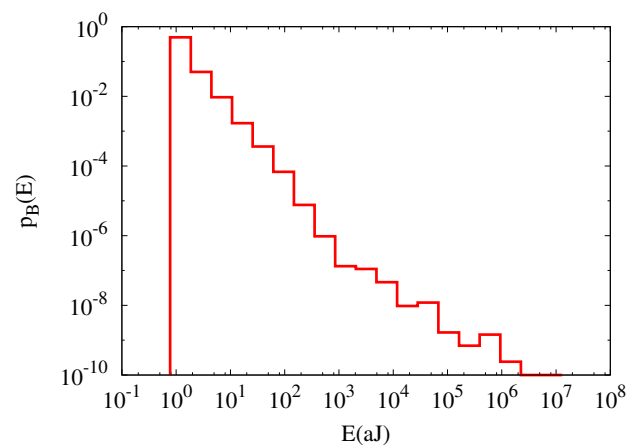,width=6cm,angle=270,clip=}
\end{center} 
\caption{\label{FIG9} Energy distribution of the AE events for a case with 45.8\% porosity (2A50R). The histogram corresponds to the signals registered in the loading regime.}
 \end{figure}

\begin{figure}[htb] 
\begin{center}
\epsfig{file=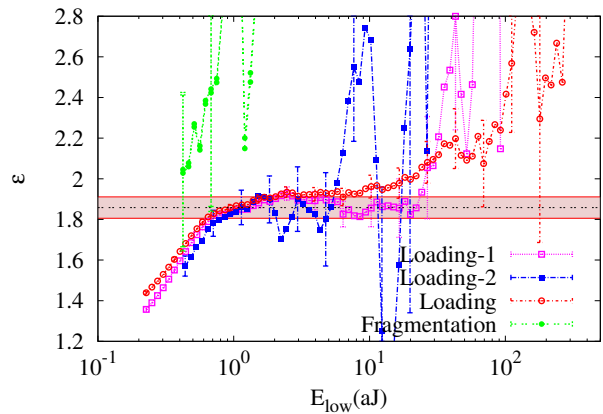,width=6cm,angle=270,clip=}
\end{center} 
\caption{\label{FIG10} Maximum likelihood fitted exponent ($\epsilon$) as a function of the lower cutoff  for data recorded during the regimes indicated in the legend and explained in the text. Data corresponds to an experiment with a 45.8\% porosity specimen.  The higher cutoff for the ML fit has been fixed to $10^3$ aJ. The horizontal band shows the estimated exponent and error bar.}
 \end{figure}

\begin{figure}[htb] 
\begin{center}
\epsfig{file=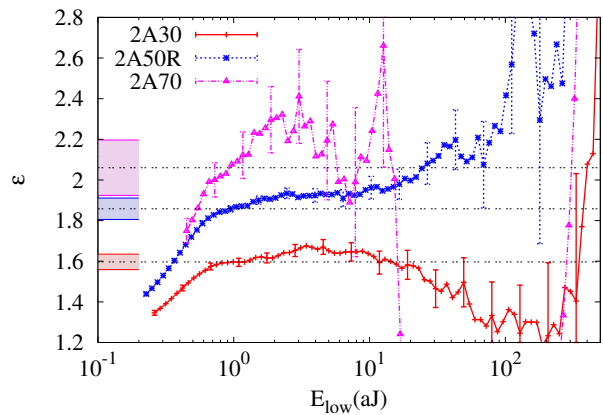,width=6cm,angle=270,clip=}
\end{center} 
\caption{\label{FIG11} Maximum likelihood fitted exponent (corresponding to the loading regime) as a function of the lower cutoff for different porosities as indicated.  }
 \end{figure}

In this section, we present the study of the statistical properties of the energy of individual acoustic signals. Fig.~\ref{FIG9} shows an example of the distribution of energies $p(E)$ corresponding to 45.8\% porosity. The data corresponds to AE events recorded during the loading regime. The distributions follow approximate power laws over some 6 decades with extended tails (see Fig.~\ref{FIG9}).  In order to examine in more detail whether or not the distribution tails decay like a power-law, we apply the method presented in Refs.~\cite{Clauset2009} and \cite{Baro2012}. The technique consists of studying the behavior of the power-law exponent ($\epsilon$) fitted with maximum likelihood method as function of a lower and higher cutoff of the data. A true power-law behavior can then be identified as a plateau (see Figs.~\ref{FIG10} and \ref{FIG11}) if $\epsilon$ is stable within error bars for several decades.

Fig.~\ref{FIG10} shows an example of the behaviour of the exponent $\epsilon$ as a function of the lower cutoff  for a higher cutoff of $10^3$ aJ for a $\Phi$=45.8\% specimen. The different curves correspond to the analysis of the data in the loading and fragmentation regimes. We also show the behaviour of the data in two subsets of the loading regime: loading-1 corresponds to the loading previous to the big crash and loading-2 to the data acquired during the big crash.   As can be seen a flat plateau can be observed  during the full loading regime, and also for the two studied subsets 1 and 2. Contrarily, the fragmentation regime does not show a clear plateau.  

Fig.~\ref{FIG11} shows a similar analysis of the fitted exponent (in the loading regime) as a function of porosity. The plateau can be clearly identified for $\Phi$=45.8\% porosity. For the ther cases,  although the plateau is not as extended, by averaging the curves between 0.5 aJ and 100 aJ, we obtain an effective exponent ($\epsilon_{eff}$) which shows a tendency to decrease with increasing porosity.

\begin{figure} [htb]
\begin{center}
\epsfig{file=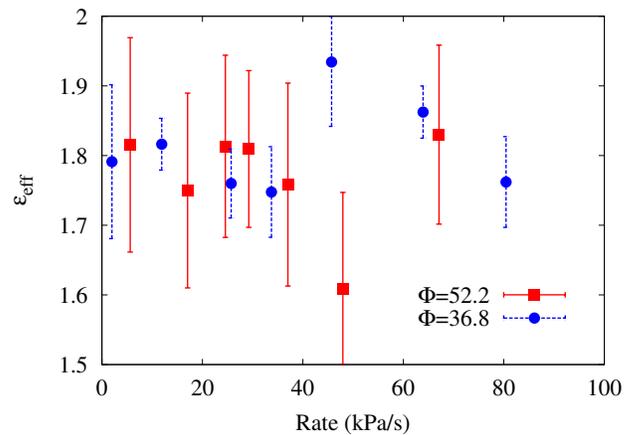,width=6cm,angle=270,clip=}
\end{center} 
\caption{\label{FIG12} Dependence of the fitted energy exponent ($\epsilon_ {eff}$) as a function of the compression rate for samples with $\Phi$=52.2\% and $\Phi$=36.8\% porosities.}
 \end{figure}

Fig.~\ref{FIG12} presents the energy exponents  for $\Phi$=52.2\% and $\Phi$=38.8\% as function of the compression rate. No systematic dependence can be found within our experimental resolution.  

\begin{figure}[htb]
\begin{center}
\epsfig{file=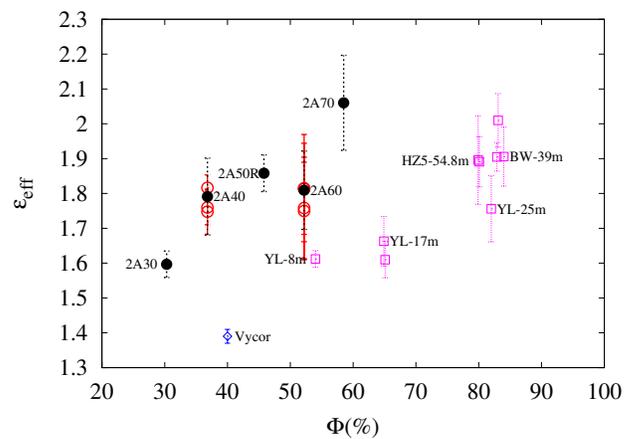,width=6cm,angle=270,clip=}
\end{center} 
\caption{\label{FIG13} Dependence of the effective exponent $\epsilon_{eff}$ with the porosity $\Phi$. Labels starting with "2A" correspond to this work. Label "Vycor" corresponds to data from Ref. \cite{Salje2011a} and the rest of labels correspond to Goethite data in Ref.~\cite{Salje2013}}
 \end{figure}

Fig.~\ref{FIG13} shows a compilation of the fitted effective exponents in the loading regimes as  function of porosity. Previous data from goethite and Vycor are also shown for comparison. On the one hand, we can observe that the behaviour of the alumina data shows a clear tendency for the effective exponent to increase with porosity, in agreement with previous results in goethite\cite{Salje2013}. Nevetheless, the overall picture suggests that data concentrates around a value of the exponent $\simeq 1.8$ for both goethite and alumnia.

The quality of the power-law fit (range of the measured plateau in Fig. \ref{FIG11}) for intermediate values of the porosity ($\Phi$=45.8\%) is much better than for much higher or lower porosities. This is an indication that we could encounter a tuned criticality scenario in porous Al$_2$O$_3$: for low porosities the system appears supercritical (rendering a small $\epsilon_{eff}$) while for high porosities the system	 appears subcritical	(large	$\epsilon_{eff}$).	Only in the intermediate range 40\% - 50\% an extended  power-law behaviour with absence of characteristic scales would occur.  The critical value of  $\Phi$ would be system dependent and, for goethite it will be higher than for alumina.

\section{Summary and conclusions}

We have performed systematic studies of the AE during the uniaxial compression on alumina samples with different porosities. The samples have been characterized by XRD diffraction and SEM. Samples in the form of  parallelepipeds were prepared into different heights and compressed with at rates between   5k Pa/s and 80 kPa/s.
The compression process is dominated by  a major failure event with a  change in height of more than 50\%. The signals recorded during the loading regime (including the big crash)  exhibit similar stationary statistical properties. The energy distributions of the AE signals were studied by using maximum likelihood techniques. The results show essentially a power law distribution, at least for porosities in the range $\Phi$=40\%-50\%. The energy range of the power law distribution depends on the porosity and we can speculate that AE during compression of porous Al$_2$O$_3$ exhibit criticality in the loading regime. Criticality would be observed only for a certain value of the porosity ($\Phi_c$). For lower porosities $\Phi < \Phi_c$ the behavior is  supercritical (rendering a small effective exponent), with less acoustic small events and big events associated with the failure. For higher porosities $\Phi > \Phi_c$ the behavior is subcritical with an exponential cutoff (rendering a higher effective exponent). This behaviour will be compatible with what has been observed in natural goethite. Both  materials will share the same value of the critical exponent $\epsilon=1.8$ and different values of $\Phi_c$. On the other hand, the existing data for SiO$_2$ would correspond to a different universality class, with a value of the exponent which is clearly lower. More studies are needed in order to identify the reasons behind these differences.

\ack

We acknowledge financial support from the Spanish Ministry of Science (MAT2010-15114). E.K.H.S. thanks the Leverhulme foundation (RG66640) and EPSRC (RG66344) for financial support. P.C.-V. acknowledges support from CONACYT (M\'exico) under scholarship No 186474. P.S. and W.M.K. acknowledge a United States Army Research Office MURI grant (W911NF-09-1-0436). The scanning electron microscopy (SEM) work was carried out in the Frederick Seitz Materials Research Laboratory at the University of Illinois at Urbana-Champaign.

\end{document}